\documentclass[aps,prl,twocolumn,superscriptaddress,showpacs,letterpaper]{revtex4}

\usepackage{graphicx}
\usepackage{amsmath, amsthm, amssymb}

\usepackage{setspace}

\newcommand{\X}{$X(3872)$}

\begin{document}

\title{Analysis of the Quantum Numbers $J^{PC}$ of the $X(3872)$ Particle}


\affiliation{Institute of Physics, Academia Sinica, Taipei, Taiwan 11529, Republic of China} 
\affiliation{Argonne National Laboratory, Argonne, Illinois 60439} 
\affiliation{Institut de Fisica d'Altes Energies, Universitat Autonoma de Barcelona, E-08193, Bellaterra (Barcelona), Spain} 
\affiliation{Baylor University, Waco, Texas  76798} 
\affiliation{Istituto Nazionale di Fisica Nucleare, University of Bologna, I-40127 Bologna, Italy} 
\affiliation{Brandeis University, Waltham, Massachusetts 02254} 
\affiliation{University of California, Davis, Davis, California  95616} 
\affiliation{University of California, Los Angeles, Los Angeles, California  90024} 
\affiliation{University of California, San Diego, La Jolla, California  92093} 
\affiliation{University of California, Santa Barbara, Santa Barbara, California 93106} 
\affiliation{Instituto de Fisica de Cantabria, CSIC-University of Cantabria, 39005 Santander, Spain} 
\affiliation{Carnegie Mellon University, Pittsburgh, PA  15213} 
\affiliation{Enrico Fermi Institute, University of Chicago, Chicago, Illinois 60637} 
\affiliation{Comenius University, 842 48 Bratislava, Slovakia; Institute of Experimental Physics, 040 01 Kosice, Slovakia} 
\affiliation{Joint Institute for Nuclear Research, RU-141980 Dubna, Russia} 
\affiliation{Duke University, Durham, North Carolina  27708} 
\affiliation{Fermi National Accelerator Laboratory, Batavia, Illinois 60510} 
\affiliation{University of Florida, Gainesville, Florida  32611} 
\affiliation{Laboratori Nazionali di Frascati, Istituto Nazionale di Fisica Nucleare, I-00044 Frascati, Italy} 
\affiliation{University of Geneva, CH-1211 Geneva 4, Switzerland} 
\affiliation{Glasgow University, Glasgow G12 8QQ, United Kingdom} 
\affiliation{Harvard University, Cambridge, Massachusetts 02138} 
\affiliation{Division of High Energy Physics, Department of Physics, University of Helsinki and Helsinki Institute of Physics, FIN-00014, Helsinki, Finland} 
\affiliation{University of Illinois, Urbana, Illinois 61801} 
\affiliation{The Johns Hopkins University, Baltimore, Maryland 21218} 
\affiliation{Institut f\"{u}r Experimentelle Kernphysik, Universit\"{a}t Karlsruhe, 76128 Karlsruhe, Germany} 
\affiliation{High Energy Accelerator Research Organization (KEK), Tsukuba, Ibaraki 305, Japan} 
\affiliation{Center for High Energy Physics: Kyungpook National University, Taegu 702-701, Korea; Seoul National University, Seoul 151-742, Korea; and SungKyunKwan University, Suwon 440-746, Korea} 
\affiliation{Ernest Orlando Lawrence Berkeley National Laboratory, Berkeley, California 94720} 
\affiliation{University of Liverpool, Liverpool L69 7ZE, United Kingdom} 
\affiliation{University College London, London WC1E 6BT, United Kingdom} 
\affiliation{Centro de Investigaciones Energeticas Medioambientales y Tecnologicas, E-28040 Madrid, Spain} 
\affiliation{Massachusetts Institute of Technology, Cambridge, Massachusetts  02139} 
\affiliation{Institute of Particle Physics: McGill University, Montr\'{e}al, Canada H3A~2T8; and University of Toronto, Toronto, Canada M5S~1A7} 
\affiliation{University of Michigan, Ann Arbor, Michigan 48109} 
\affiliation{Michigan State University, East Lansing, Michigan  48824} 
\affiliation{Institution for Theoretical and Experimental Physics, ITEP, Moscow 117259, Russia} 
\affiliation{University of New Mexico, Albuquerque, New Mexico 87131} 
\affiliation{Northwestern University, Evanston, Illinois  60208} 
\affiliation{The Ohio State University, Columbus, Ohio  43210} 
\affiliation{Okayama University, Okayama 700-8530, Japan} 
\affiliation{Osaka City University, Osaka 588, Japan} 
\affiliation{University of Oxford, Oxford OX1 3RH, United Kingdom} 
\affiliation{University of Padova, Istituto Nazionale di Fisica Nucleare, Sezione di Padova-Trento, I-35131 Padova, Italy} 
\affiliation{LPNHE, Universite Pierre et Marie Curie/IN2P3-CNRS, UMR7585, Paris, F-75252 France} 
\affiliation{University of Pennsylvania, Philadelphia, Pennsylvania 19104} 
\affiliation{Istituto Nazionale di Fisica Nucleare Pisa, Universities of Pisa, Siena and Scuola Normale Superiore, I-56127 Pisa, Italy} 
\affiliation{University of Pittsburgh, Pittsburgh, Pennsylvania 15260} 
\affiliation{Purdue University, West Lafayette, Indiana 47907} 
\affiliation{University of Rochester, Rochester, New York 14627} 
\affiliation{The Rockefeller University, New York, New York 10021} 
\affiliation{Istituto Nazionale di Fisica Nucleare, Sezione di Roma 1, University of Rome ``La Sapienza," I-00185 Roma, Italy} 
\affiliation{Rutgers University, Piscataway, New Jersey 08855} 
\affiliation{Texas A\&M University, College Station, Texas 77843} 
\affiliation{Istituto Nazionale di Fisica Nucleare, University of Trieste/\ Udine, Italy} 
\affiliation{University of Tsukuba, Tsukuba, Ibaraki 305, Japan} 
\affiliation{Tufts University, Medford, Massachusetts 02155} 
\affiliation{Waseda University, Tokyo 169, Japan} 
\affiliation{Wayne State University, Detroit, Michigan  48201} 
\affiliation{University of Wisconsin, Madison, Wisconsin 53706} 
\affiliation{Yale University, New Haven, Connecticut 06520} 
\author{A.~Abulencia}
\affiliation{University of Illinois, Urbana, Illinois 61801}
\author{J.~Adelman}
\affiliation{Enrico Fermi Institute, University of Chicago, Chicago, Illinois 60637}
\author{T.~Affolder}
\affiliation{University of California, Santa Barbara, Santa Barbara, California 93106}
\author{T.~Akimoto}
\affiliation{University of Tsukuba, Tsukuba, Ibaraki 305, Japan}
\author{M.G.~Albrow}
\affiliation{Fermi National Accelerator Laboratory, Batavia, Illinois 60510}
\author{D.~Ambrose}
\affiliation{Fermi National Accelerator Laboratory, Batavia, Illinois 60510}
\author{S.~Amerio}
\affiliation{University of Padova, Istituto Nazionale di Fisica Nucleare, Sezione di Padova-Trento, I-35131 Padova, Italy}
\author{D.~Amidei}
\affiliation{University of Michigan, Ann Arbor, Michigan 48109}
\author{A.~Anastassov}
\affiliation{Rutgers University, Piscataway, New Jersey 08855}
\author{K.~Anikeev}
\affiliation{Fermi National Accelerator Laboratory, Batavia, Illinois 60510}
% ...
\collaboration{CDF Collaboration\footnote{With visitors from $^a$University of Athens, 
$^b$University of Bristol, 
$^c$University Libre de Bruxelles, 
$^d$Cornell University, 
$^e$University of Cyprus, 
$^f$University of Dublin, 
$^g$University of Edinburgh, 
$^h$University of Heidelberg, 
$^i$Universidad Iberoamericana, 
$^j$University of Manchester, 
$^k$Nagasaki Institute of Applied Science, 
$^l$University de Oviedo, 
$^m$University of London, Queen Mary and Westfield College, 
$^n$Texas Tech University, 
$^o$IFIC(CSIC-Universitat de Valencia), 
}}
\noaffiliation

\begin{abstract}
We present an analysis of angular distributions and correlations
of the \X{} particle in the exclusive decay mode $X(3872) \to J/ \psi \, \pi^+ \pi^-$
with $J/ \psi \to \mu^+ \mu^-$. We use 780 pb$^{-1}$ of data from
$p\bar{p}$ collisions at $\sqrt{s} = 1.96$ TeV collected with
the CDF~II detector at the Fermilab Tevatron.
We derive constraints on spin, parity, and charge conjugation
parity of the \X{} particle
by comparing measured angular distributions of the decay
products with predictions for different $J^{PC}$ hypotheses.
The assignments $J^{PC} = 1^{++}$ and $2^{-+}$ are the only
ones consistent with the data.
\end{abstract}

\maketitle

The recent discovery of the \X{} particle
\cite{X3872:Belle,X3872:CDF} has revived general interest
in charmonium spectroscopy.
The exact nature of this particle is still unknown.
Attempts to explain the \X{} particle as a conventional bound
quark-antiquark state have shortcomings, such as deviations from
mass predictions
or violation of isospin conservation \cite{Barnes:2003vb}.
The close proximity of the \X{} particle mass to the $D^0\bar{D}^{\ast 0}$
mass threshold has raised the question whether the $X(3872)$ particle is
an exotic form of matter \cite{Barnes:2003vb}.
The determination of the quantum numbers spin $J$, parity $P$,
and charge conjugation parity~$C$ is of
vital importance for establishing the nature of the \X{} particle.
The evidence for the decay mode $X(3872) \to J/\psi \; \gamma$ \cite{Abe:2005ix}
and the measurement of the dipion mass distribution \cite{Abulencia:2005zc},
which is in agreement with the decay mode $X(3872) \to J/\psi \rho^0$,
are consistent with a $C$-even assignment.
Reference \cite{Gokhroo:2006bt} observes an enhancement in
the $D^0 \bar{D}^0\pi^0$ mass spectrum and concludes that, if
assigned to the \X{} particle, low values for the spin quantum number are
favored.
Neglecting effects from model uncertainties in the dipion mass spectrum
(see \cite{Abulencia:2005zc}), preliminary results
from \cite{Abe:2005iy} favor $J^{PC} = 1^{++}$.
In this Letter we report the angular distributions
in the decay
$X(3872) \to J/ \psi \, \pi^+ \pi^-$, $J/\psi\to \mu^+\mu^-$,
and compare them with predictions for different $J^{PC}$ states.
The analysis is independent of any specific model of the
internal structure of the \X{} particle. We consider all allowed
states up to spin two and $C$-odd spin three states.


We use a sample of $p \bar{p}$ collisions at a center-of-mass
energy of $\sqrt{s} = 1.96$ TeV with an integrated luminosity
of 780 pb$^{-1}$ collected with the CDF II detector
at the Fermilab Tevatron.
The CDF II detector \cite{Acosta:2004yw} consists of a magnetic
spectrometer surrounded by electromagnetic and hadronic calorimeters
and muon detectors.
The tracking system is composed of a silicon micro-strip
detector \cite{Hill:2004qb}
surrounded by an open-cell drift chamber called the central
outer tracker (COT) \cite{Affolder:2003ep}.
We detect muons in planes of multiwire drift
chambers \cite{Ascoli:1987av} in the pseudorapidity range
$|\eta| \le 1.0$.
The $J/\psi \rightarrow \mu^+ \mu^-$ decays used in this
analysis are recorded using a dimuon trigger, which requires
two oppositely charged COT tracks matched to muon chamber
track segments with an invariant mass from 2.7 to 4.0 GeV/$c^2$.

\begin{figure}
\begin{center}
\includegraphics[width=8.2cm]{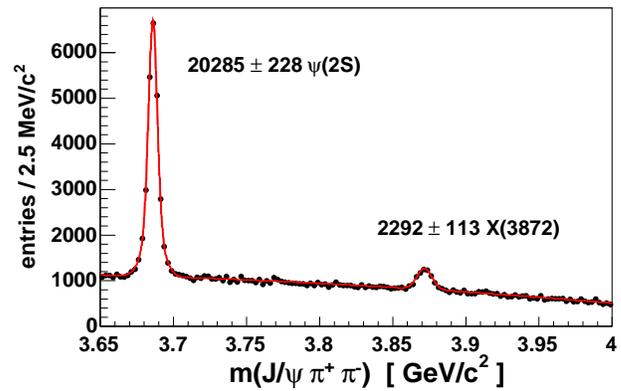}
 \caption{\label{fig:xwp}The $J/\psi\,\pi^+\pi^-$ mass spectrum after
optimizing the selection cuts, fitted by a double
Gaussian function for the $\psi(2S)$ (left), a Gaussian function 
for the \X{} particle (right), and a second order
polynomial for the combinatorial background.}
\end{center}
\end{figure}

The basic event selection is described in
\cite{X3872:CDF,Abulencia:2005zc}, although we do not cut on
the dipion mass.
Additional criteria are imposed on the number of candidates
per event, the transverse momentum $p_T$ of the \X{}
particle candidate ($>6$~GeV/$c$),
the $p_T$ of the $J/\psi$ ($>4$~GeV/$c$), and the kinetic energy
released in the \X{} particle decay,
$Q = m(J/ \psi\, \pi \pi) -
m(J/ \psi) - m(\pi \pi)$ ($<100$~MeV/$c^2$), where $m(J/ \psi)$ is
from \cite{PDBook}.
The cuts are chosen to optimize the significance $S/\sqrt{S+B}$
of the observed signal,
where $S$ and $B$ are the fitted number of signal and combinatorial
background
events in a $\pm1.5\sigma$ window centered on the \X{} particle mass.
The resulting distribution of the invariant
$J/\psi\,\pi^+\pi^-$ mass is shown in Fig. \ref{fig:xwp}.

To simulate the decays of \X{} particle states with specific $J^{PC}$
assumptions, we first generate phase-space decays of
$X(3872)\to J/\psi\, \pi^+\pi^-$, $J/\psi \to \mu^+\mu^-$.
Detector effects are included
using parameterized efficiencies and acceptances. This sample is
weighted according to each specific $J^{PC}$ hypothesis using the
corresponding matrix element
$\mathcal{M}_{tot}$ described below.

The decay of the narrow \X{} particle is modeled as the sequential two-body decay
chain $\X{} \to J/\psi\; (\pi^+\pi^-)$, $J/\psi\to\mu^+\mu^-$
and the decay of the intermediate $(\pi^+\pi^-)$ state to 
$\pi^+\pi^-$. Assuming low relative angular momentum between
the pions and conservation of $C$ parity, the
intermediate pion state can be
in either a relative $S$-wave ($(\pi\pi)_S$)
or a $P$-wave ($\rho^0$) state.
$\mathcal{M}_{tot}$ is formed by the product of a matrix
element $\mathcal{M}_i$ for each decay and a term $T(m({\pi\pi}))$,
which describes the mass dependence of the
intermediate $(\pi^+\pi^-)$ system.
Because of the very narrow width of the intermediate $J/\psi$, we
can neglect the $J/\psi$ mass dependence.

With fixed helicities the angular dependence of
a two-body decay amplitude is given by
the Wigner function
$D^{J_{i}}_{\lambda_{i},\lambda_{i,1}-\lambda_{i,2}}$
\cite{Chung:1997jn,Varshalovich:1988ye},
where $J_{i}$ and $\lambda_{i}$
are the spin and helicity of the
decaying particle, and $\lambda_{i,1}$ and $\lambda_{i,2}$ are the
helicities of the child particles in the parent rest frame.
The function is multiplied by two Clebsch-Gordan
coefficients, coupling the spins of the child particles to
their summed spin $S_i$,
and $S_i$ with their relative angular momentum $L_i$ to $J_i$.

In general, in the $\X \to J/\psi\; (\pi^+\pi^-)$ decay there
is more than one combination
to form $J$ from $L$ and $S$ in a parity-conserving way.
Of the independent amplitudes corresponding to these combinations,
only the ones with lowest $L$, assumed to be dominant, are taken into
account.
If more than one amplitude remains,
mixing parameters are introduced to describe the physical state.
Since the virtual photon in the $J/\psi \to \mu^+\mu^-$ decay
can be treated as transverse, helicity combinations
with $\lambda_{\mu^+}-\lambda_{\mu^-}=0$ are neglected.

The dependence of $\mathcal{M}_{tot}$ on the dipion mass has
model ambiguities. Therefore, we do not use the information from
$m(\pi\pi)$ to distinguish between different $J^{PC}$ hypotheses.
The influence of the $m(\pi\pi)$ model
on the angular distributions via acceptance effects is very small.
Nevertheless we choose for all $J^{PC}$ hypotheses the same model for
the $m(\pi\pi)$-dependent terms, which agrees with the $m(\pi\pi)$
spectrum measurement. In this way, no hypothesis is rejected due
to a wrong $m(\pi\pi)$ model.
In detail, we fix $T(m({\pi\pi}))$ to a relativistic Breit-Wigner formula
with mass and width of a $\rho^0$ \cite{PDBook}.
Following \cite{Abulencia:2005zc}, we also fix the momentum dependence of the
matrix element of the $(\pi^+\pi^-) \to \pi^+ \pi^-$ decay
to ${k^{\ast}} \cdot f_1\left(k^{\ast}\right)$, where $k^{\ast}$ is
the magnitude of the three-momentum of one of the pions
in the $(\pi^+\pi^-)$ rest frame and $f_1\left(k^{\ast}\right)$ is
a Blatt-Weisskopf form factor \cite{BlattWeisskopf} to counter the
divergence for rising $k^{\ast}$. This form factor has
the effective size $r$ of the particle as a free parameter
which we set to a common choice of $r=1$ fm.

A weight is formed from the square of the total matrix element
$\mathcal{M}_{tot}$ by
averaging over all initial state helicities assuming unpolarized
\X{} particle production, incoherently summing
over all final state helicities, and coherently summing over all
intermediate state helicities.

\begin{figure*}
\begin{center}
\includegraphics[height=3.5cm]{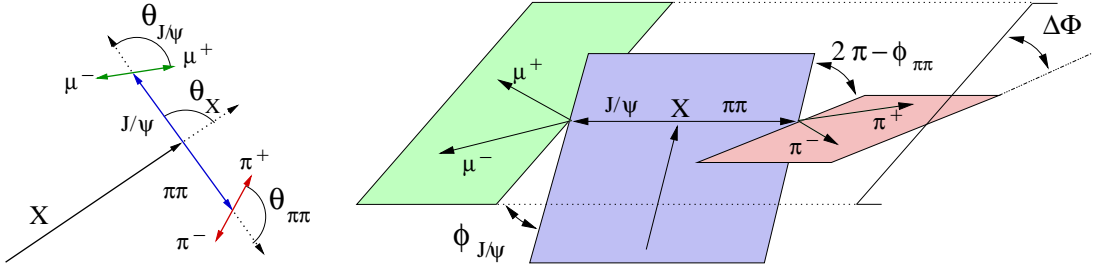}
\caption{Definition of the decay angles. The polar
angles ($\theta$) are calculated from the parent momenta and the child
momenta in the corresponding parent rest frame.}
\label{fig:def_angles}
\end{center}
\end{figure*}

The decay is described by the decay angles 
$\theta_X$, $\theta_{J/ \psi}$, $\phi_{J/ \psi}$, 
$\theta_{\pi\pi}$, $\phi_{\pi\pi}$, and $\Delta\Phi$,
as defined in Fig.~\ref{fig:def_angles}.
For unpolarized \X{} particle production and because of rotational symmetry, the
$J^{PC}$ of the \X{} particle and the $(\pi^+\pi^-)$ system affect the distribution
of only four variables:
$m(\pi\pi)$, $\cos(\theta_{J/ \psi})$, $\cos(\theta_{\pi\pi})$, and
$\Delta\Phi$.

The angular distributions are analyzed with a three-dimensional fit
to take into account their correlations. From simulation studies, the
optimal binning is determined to be three bins in
$||\Delta\Phi-\pi|-\frac{\pi}{2}|$, and two bins in each of
$|\cos(\theta_{J/\psi})|$ and $|\cos(\theta_{\pi\pi})|$, where
absolute values are used to exploit final state charge symmetry.
The invariant $J/ \psi \, \pi^+ \pi^-$ mass spectrum
is fitted in each of the resulting 12 bins in a mass window of
$\pm110$~MeV/$c^2$ around the $X(3872)$ particle position
using a binned maximum likelihood fit,
where the bin width is 2.5 MeV/$c^2$.
The distribution is described by a Gaussian function
for the $X(3872)$ particle
and a second order polynomial for the background.
The position and width
of the Gaussian function describing the
\X{} particle are first determined from a fit
to the full invariant mass spectrum and are then fixed
in the subsequent fits. We compare the fitted yield as a function
of the angular variables
with the predictions for different $J^{PC}$ assignments
by forming a $\chi^2$ based on statistical uncertainties of the
measurement.
We determine the normalization of the simulated
distributions from the measurement so that 11 degrees
of freedom remain.

The decay amplitude for the state with
$J^{PC} = 1^{-+}$ consists of
three $LS$-terms with the same $L$ value; the $J^{PC} = 2^{-+}$ state
has an amplitude with two $LS$-terms (see Table \ref{tab:XResult}).
None of the $1^{-+}$ terms describes the data alone, so
we fit for a mixed state by minimizing the $\chi^2$.
For the $2^{-+}$ state, the amplitude for $S=1$ is sufficient to
describe the data.

\begin{table}
\begin{center}
\begin{tabular}{ccccc}
\hline \hline
$J^{PC}$ & decay & $LS$ &   $\chi^2\;(11$ d.o.f.)& $\chi^2$ prob.\\
\hline
$1^{++}$  &$J/\psi \rho^0$     & 01       &    13.2         &   0.28 \\
$2^{-+}$  &$J/\psi \rho^0$     & 11,12    &    13.6         &   0.26 \\
$1^{--}$  &$J/\psi (\pi\pi)_S$ & 01       &    35.1         &   2.4$\times10^{-4}$     \\
$2^{+-}$  &$J/\psi (\pi\pi)_S$ & 11       &    38.9         &   5.5$\times10^{-5}$     \\
$1^{+-}$  &$J/\psi (\pi\pi)_S$ & 11       &    39.8         &   3.8$\times10^{-5}$     \\
$2^{--}$  &$J/\psi (\pi\pi)_S$ & 21       &    39.8         &   3.8$\times10^{-5}$     \\
$3^{+-}$  &$J/\psi (\pi\pi)_S$ & 31       &    39.8         &   3.8$\times10^{-5}$     \\
$3^{--}$  &$J/\psi (\pi\pi)_S$ & 21       &    41.0         &   2.4$\times10^{-5}$     \\
$2^{++}$  &$J/\psi \rho^0$     & 02       &    43.0         &   1.1$\times10^{-5}$     \\
$1^{-+}$  &$J/\psi \rho^0$     & 10,11,12 &    45.4         &   4.1$\times10^{-6}$     \\
$0^{-+}$  &$J/\psi \rho^0$     & 11       &     104         &   3.5$\times10^{-17}$    \\
$0^{+-}$  &$J/\psi (\pi\pi)_S$ & 11       &     129        &   $\le$1$\times10^{-20}$ \\
$0^{++}$  &$J/\psi \rho^0$     & 00       &     163         &   $\le$1$\times10^{-20}$ \\
\hline
\hline
\end{tabular}
\end{center}
\caption{\label{tab:XResult}Result of the $X(3872)$ particle angular analysis.
Listed are the state, the decay mode, the $L$ and $S$ quantum numbers
of the $J/\psi$-$(\pi^+\pi^-)$ system, the $\chi^2$ with 11 degrees
of freedom and the $\chi^2$ probability.}
\end{table}

\begin{figure}
\begin{center}
\includegraphics[width=8.2cm]{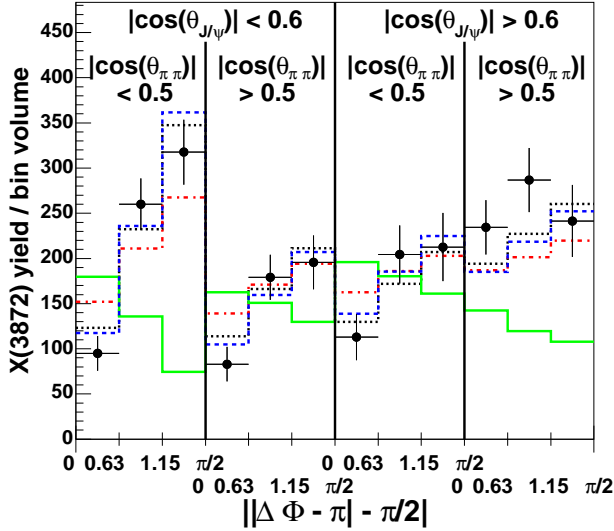}
\caption{Measured 3D angular distribution with acceptance
corrected predictions
for $J^{PC} = 0^{++}$ (solid line), $1^{++}$ (dotted line),
$2^{-+}$ (dashed line), and $1^{--}$ (dash-dotted line). The plot is divided into
2$\times$2 regions, corresponding to intervals of
$|\cos(\theta_{J/\psi})|$ and $|\cos(\theta_{\pi\pi})|$. Each region
shows the distribution of $||\Delta\Phi-\pi|-\frac{\pi}{2}|$ in 3 bins.
The bin contents have been scaled to the same bin volume.}
\label{fig:3d_dist}
\end{center}
\end{figure}

Table \ref{tab:XResult} shows the $\chi^2$
for each $J^{PC}$ assignment.
We find that only the assignments $J^{PC} = 1^{++}$
and $2^{-+}$ are able to describe the data. All other states are
rejected by more than 3 standard deviations
($\chi^2$ prob. $\le2.7 \times 10^{-3}$).
Figure \ref{fig:3d_dist} shows the measurement and the 
expected distribution for four of the assignments.

An important cross-check of the analysis is
to verify whether the correct result is obtained
for the
$\psi(2S)$, with known quantum numbers $J^{PC} = 1^{--}$,
which decays into the same exclusive final state as the \X{} particle.
For the $1^{--}$ assignment, the fit probability is $1.5\%$.
Using the $\psi(2S)$ model of Novikov and
Shifman \cite{Novikov:1980fa}, which includes a 
small $D$-wave admixture in the description
of the $(\pi^+\pi^-)$ system, the fit probability is $17.9\%$.
The sensitivity to such a small admixture is only present in the high
statistics $\psi(2S)$ sample.
The next best model
$J^{PC} = 2^{++}$ has a fit probability of $0.58\%$, and
all other hypotheses that were tested yielded fit probabilities smaller
than $2\times 10^{-6}$.

\begin{figure}
\begin{center}
\includegraphics[width=8.2cm]{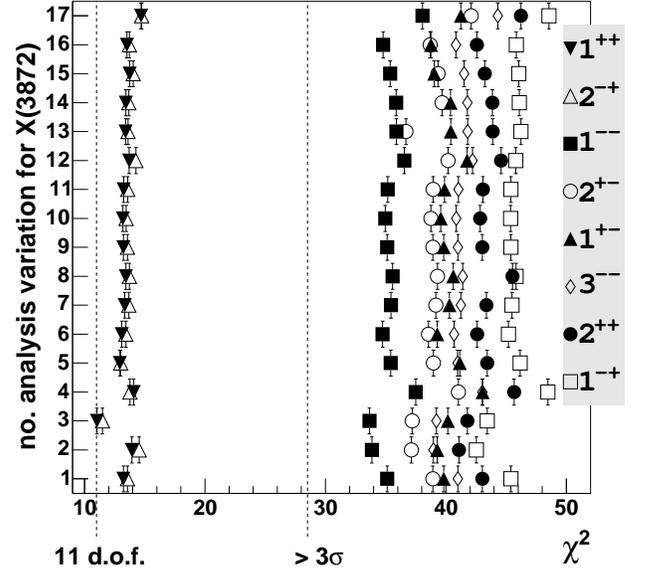}
\end{center}
\caption{Total $\chi^2$ for different
analysis variations on the y-axis, explained in the text.
Vertical bars are added for visual guidance.
The $\chi^2$ values of the spin 0 states are all above 100.
The $2^{--}$ and $3^{+-}$ states have the same angular distribution
as the $1^{+-}$ state.}
\label{fig:syst_graph}
\end{figure}

We vary several inputs to the fitting procedure
and the model of the $X(3872)$ particle to
investigate the stability of the $\chi^2$. Figure \ref{fig:syst_graph}
shows the resulting $\chi^2$ values for the different $J^{PC}$
hypotheses for the variations investigated. The default analysis is shown
as variation (1).
The following effects are considered:
(2)/(3) decrease/increase the fit window by 20 MeV/$c^2$,
(4)/(5) decrease/increase the bin width to 2.0/2.86 MeV/$c^2$,
(6)/(7) vary fixed $X(3872)$ particle mass by $\pm 1 \sigma$,
(8)/(9) vary fixed $X(3872)$ particle width by $\pm 1 \sigma$.

To evaluate the contribution to the systematic uncertainty from
our choice of the $m(\pi\pi)$ spectrum, the
following variations are considered:
(10) fix form-factor $r$ to 0.001 fm,
(11) fix form-factor $r$ to 100.0 fm,
(12) use simple phase space for $m(\pi\pi)$. 

Finally, systematic uncertainty due to details concerning
the simulation has been considered by varying distributions for
(13) $p_T$ and
(14) $\eta$ of the $X(3872)$ particle,
switching off
(15) a $p_T$ dependent efficiency correction for the pions,
(16) a $\phi$ dependent correction of the COT, and
(17) an effective $\eta$ correction used to model the position
of the generated primary vertex.
All variations are consistent with $1^{++}$ and $2^{-+}$ being the
only likely assignments.


A conventional explanation for the \X{} resonance is
a charmonium ($c\bar{c}$) state.
In this picture, the state with $J^{PC}=1^{++}$
could be identified with the $\chi_{c1}^\prime$
and the assignment $J^{PC} = 2^{-+}$ with the $\eta_{c2}$.
An exotic interpretation is that the \X{} particle
is a molecular state or that a significant four-quark
interaction contributes to the wave-function \cite{DeRujula:1976qd}.
The result of this analysis
is compatible with the models of a molecular state
developed by Tornqvist \cite{Tornqvist:2004qy} and
Swanson \cite{Swanson:2003tb},
who predict the quantum numbers $J^{PC}=1^{++}$
for a bound $D\bar{D}^{*}$ state.


In summary, a spin-parity analysis of the $X(3872)$ particle
in the final state $\mu^+\mu^-\,\pi^+\pi^-$ has been performed.
The method of helicity amplitudes has been used to analyze
$X(3872) \to J/\psi (\pi\pi)_S$ and $X(3872) \to J/\psi \rho^0$ transitions.
Using a $\chi^2$ approach to compare expected angular distributions
with measured distributions, it
is found that only the $C$-even assignments $J^{PC} = 1^{++}$
and $2^{-+}$, both decaying via $J/\psi \rho^0$, describe the data.
All other states are excluded at 99.7\% confidence level.


We thank the Fermilab staff and the technical staffs of the participating institutions for their vital contributions. This work was supported by the U.S. Department of Energy and National Science Foundation; the Italian Istituto Nazionale di Fisica Nucleare; the Ministry of Education, Culture, Sports, Science and Technology of Japan; the Natural Sciences and Engineering Research Council of Canada; the National Science Council of the Republic of China; the Swiss National Science Foundation; the A.P. Sloan Foundation; the Bundesministerium f\"ur Bildung und Forschung, Germany; the Korean Science and Engineering Foundation and the Korean Research Foundation; the Particle Physics and Astronomy Research Council and the Royal Society, UK; the Institut National de Physique Nucleaire et Physique des Particules/CNRS; the Russian Foundation for Basic Research; the Comisi\'on Interministerial de Ciencia y Tecnolog\'{\i}a, Spain; the European Community's Human Potential Programme under contract HPRN-CT-2002-00292; and the Academy of Finland.


\end{document}